\documentclass[aps,prb,twocolumn,showpacs,superscriptaddress, longbibliography]{revtex4-2}

\usepackage{xcolor, graphicx}
\usepackage{amssymb,amsmath, amsfonts, bm}
\usepackage{newtxtext,newtxmath}
\usepackage{float}
\usepackage{dcolumn}
\usepackage[none]{hyphenat} 
\usepackage{tabularx}

\usepackage{multirow}
\usepackage{hyperref}
\hypersetup{
    colorlinks=true,
    linkcolor=blue,    
    urlcolor=blue,
    citecolor=blue,
    }
    
\usepackage{multibib}
\newcites{SI}{Reference}
\usepackage{easyReview}

\begin{document}
\title{Correlated flat bands in the paramagnetic phase of triangular antiferromagnets Na$_2$BaX(PO$_4$)$_2$ (X = Mn, Co, Ni)}

\author{Cong Hu}
 \affiliation{School of Physical Science and Technology, ShanghaiTech University, Shanghai 201210, China}
\author{Xuefeng Zhang}
 \affiliation{School of Physical Science and Technology, ShanghaiTech University, Shanghai 201210, China}
\author{Yunlong Su}
 \affiliation{School of Physical Science and Technology, ShanghaiTech University, Shanghai 201210, China}
\author{Gang Li}
 \email{ligang@shanghaitech.edu.cn}
\affiliation{School of Physical Science and Technology, ShanghaiTech University, Shanghai 201210, China}
\affiliation{\mbox{ShanghaiTech Laboratory for Topological Physics, ShanghaiTech University, Shanghai 201210, China}}

\date{\today}

\begin{abstract}
Flat band systems in condensed matter physics are intriguing because they can exhibit exotic phases and unconventional properties. In this work, we studied three correlated magnetic systems, Na$_2$BaX(PO$_4$)$_{2}$ (X = Mn, Co, Ni), and revealed their unusual electronic structure and magnetic properties. Despite their different effective angular momentum, our first-principles calculations showed a similar electronic structure among them. However, their different valence configurations led to different responses to electronic correlations in the high-temperature paramagnetic phase. Using the dynamical mean-field method, we found that all systems can be understood as a multi-band Hubbard model with Hund's coupling. Our calculations of spin susceptibility and the {\it ab-initio} estimation of magnetic exchange coupling indicated strong intra-plane antiferromagnetic coupling and weak inter-plane coupling in all systems. The ground states of these systems are largely degenerate. It is likely that none of these magnetic states would dominate over the others, leading to the possibility of quantum spin liquid states in these systems. Our work unifies the understanding of these three structurally similar systems and opens new avenues for exploring correlated flat bands with distinct electronic and magnetic responses.
\end{abstract}

\maketitle

\section{Introduction}
Flat band systems are systems with electronic bands around the Fermi level of very small bandwidth.  
Flat bands often occur in transition metal (TM) compounds, as the partially filled $d$-orbitals are localized in nature making the flat bands easier to form. 
The small bandwidth leads to an enhanced electron-electron interaction and quenched electron kinetic energy. 
The strong electronic correlation is the source of many exotic phenomena, including the spontaneous symmetry breaking to magnetic/charge density waves, non-Fermi liquid behavior, unconventional superconductivity, quantum spin-liquid states (QSL), etc. 
Thus, flat band systems are of particular potential to host these unconventional phenomena.

Flat bands in pristine solids usually appear in special lattices, such as the Kagome, pyrochlore, and Lieb lattices~\cite{kagome_lieb, Checker-board_lattice, Liu_2014}.
The peculiar arrangement of sites in these lattices renders one or more eigenmodes of electron motion with spatially localized orbitals.
The spatial extension of these orbitals may confine to a lattice site manifesting the flat bands as atomic energy levels. 
These orbitals can also span over a few sites inside a small cluster leading to the molecular energy levels. 
The latter has been discovered in LiZn$_{2}$Mo$_3$O$_8$~\cite{LiZn2Mo3O8_Sheckelton2012,LiZn2Mo3O8_FlintR2013,LiZn2Mo3O8_Mourigal2014,LiZn2Mo3O8_Sheckelton2014,LiZn2Mo3O8_Chen2016,LiZn2Mo3O8_Chen2018}, GaM$_4$X$_8$ ( M = V, Mo, Nb, Ta; X = S, Se)~\cite{GaM4X8_Yaich1984, GaM4X8_Pocha2000,GaM4X8_Abd-Elmeguid2004, GaM4X8_Pocha2005,GaM4X8_Jakob2007, GaM4X8_Sieberer2007, GaM4X8_Bichler2008, GaM4X8_Camjayi2012, GaM4X8_Malik2013,GaM4X8_PRB2020, GaM4X8_PRR2022}, and Nb$_3$Cl$_8$ \cite{Nb3Cl8_Pasco2019, PhysRevB.99.165141, PhysRevB.102.035412, PhysRevResearch.2.033001, Regmi-2022, PhysRevB.106.085418, Nb3Cl8_nanoletter, gao2022mott, Nb3Cl8_Hu2023}.

Other geometrically frustrated lattices, such as the isotropic triangular lattice, do not naturally favor flat bands. 
They are often considered as the host of QSL states and/or quantum spin state transitions (QSSTs) owing to the frustration of the various long-range orders \cite{qsl_0_lee_end_2008,qsl_1_balents_spin_2010,qsl_2_zhou_quantum_2017,qsl_3_savary_quantum_2017,qsl_4_knolle_field_2019}.
The QSL state has received substantial attention in the past 30 years because it supports non-Abelian quasiparticles \cite{nayak_non-abelian_2008} and fractional excitations, known as spinons \cite{spinon_kohno_spinons_2007,spinon_nussinov_high-dimensional_2007,spinon_han_fractionalized_2012,spinon_punk_topological_2014},
and may contribute to the development of topological quantum computation\cite{nayak_non-abelian_2008,kitaev_topological_2006}. 
Triangular compounds, including the organic salts $\kappa $-(BEDT-TTF)$_2$Cu$_2$(CN)$_3$, NaYbX$_2$(X = O, S, Se) \cite{NaYbO2_bordelon_2019, NaYbS2_baenitz_naybs_2018, NaYbS2_sarkar_quantum_2019, NaYbS2_SOC_2020, NaYbSe2_ranjith_anisotropic_2019, NaYbSe2_dai_spinon_2021, NaYbSe2_zhang_crystalline_2021}, CsYbSe$_2$ \cite{CsYbSe2_xing_field-induced_2019, CsYbSe2_xie_field-induced_2021}, YbMgGaO$_4$ \cite{YbMgGaO4_2_shen_evidence_2016, YbMgGaO4_3_li_muon_2016, YbMgGaO4_1_zhang_hierarchy_2018} and Na$_2$BaCo(PO$_4$)$_2$ \cite{Na2BaCo(PO4)2_lee_temporal_2021, Na2BaCo(PO4)2_li_possible_2020, Na2BaCo(PO4)2_zhong_strong_2019, Na2BaCo(PO4)2_mei_2022, Na2BaCo(PO4)2_liwei_2022, Co21128_J_Kataev2021, Co21128_huang2022thermal, Co21128_chen_Gang_2023} have been speculated as QSL candidates and are under intensive investigations.
Specifically, in NaYbX$_2$, experiments,  including heat capacity, magnetization, muon spectroscopy, and neutron diffraction, failed to detect any long-range order at zero field~\cite{NaYbO2_bordelon_2019, NaYbS2_baenitz_naybs_2018, NaYbS2_sarkar_quantum_2019, NaYbS2_SOC_2020, NaYbSe2_ranjith_anisotropic_2019, NaYbSe2_dai_spinon_2021, NaYbSe2_zhang_crystalline_2021}. 
CsYbSe$_2$ might be at a critical phase between the QSL and 120$^{\circ}$ magnetic order \cite{CsYbSe2_xing_field-induced_2019, CsYbSe2_xie_field-induced_2021}. 
The ground state of YbMgGaO$_4$ has also been considered as a random spin network state or a spin-liquid-like state with orientational spin disorder due to the Mg/Ga site \cite{YbMgGaO4_5_YbMgGaO4_zhu_disorder-induced_2017, YbMgGaO4_4_kimchi_valence_2018}. 
Although Na$_2$BaCo(PO$_4$)$_2$ exhibits antiferromagnetic orders at $T_N$ = 148 mK, the small residual term of the thermal conductivity shows that it acts as a gapless QSL with itinerant excitations above its $T_N$ \cite{Na2BaCo(PO4)2_li_possible_2020}.

In this work, we study the realization of flat bands in some of these isotropic triangular lattices and their spontaneous symmetry-breaking. 
Unlike the Kagome, pyrochlore, and Lieb lattices, it is much harder for triangular lattices to host flat bands. 
However, if realized, the flat bands would bring a new ingredient to the much-discussed triangular QSL states as the flat bands are more sensitive to electronic correlations and various collective instabilities. 
We focus on three particular systems Na$_2$BaX(PO$_4$)$_2$ (X = Mn, Co, Ni), and demonstrate that they realize the desired correlated flat bands with frustrated magnetic couplings. 
The purpose of our study is two folds: (1) The high-quality single-crystals of Na$_2$BaX(PO$_4$)$_2$ (X = Mn, Co, Ni) are available and there were detailed experimental studies of the three systems, but systematic theoretical understanding is still missing. 
These compounds are believed to be superior to other triangular QSL compounds as they are free of site-mixing or disorder issues. 
Signatures observed in critical experiments, such as the inelastic neutron scattering and magnetic specific heat, can be more definitely attributed to the intrinsic spinon dynamics.
(2) Na$_2$BaX(PO$_4$)$_2$ (X = Mn, Co, Ni) are rare triangular lattices with correlated flat bands. 
Furthermore, these three systems have different effective spin quantum numbers but display highly similar responses in various experiments.
Na$_2$BaCo(PO$_4$)$_2$ is an effective spin-1/2 system, but Na$_2$BaNi(PO$_4$)$_2$ and Na$_2$BaMn(PO$_4$)$_2$ has effective spin-1 and spin-5/2 . 
As compared to the much discussed spin-1/2 systems, the higher spin tends to suppress quantum fluctuations \cite{Ba3NiSb2O9_1_cheng_high-pressure_2011, Ba3NiSb2O9_2_chen_frustrated_2012, Ba3NiSb2O9_3_quilliam_gapless_2016, Ba3NiSb2O9_4_fak_evidence_2017}, thus, there are very few reports of QSL in spin-1 and spin-5/2 compounds. 
Theoretically understanding the difference of these three systems is, therefore, of great interest.

Among the three systems, Na$_{2}$Co(PO$_{4}$)$_{2}$ is unique. 
The high-spin $S=3/2$  Co$^{2+}$ ion  in the octahedral field features a predominantly $t_{2g}^{5}e^{2}_{g}$ electronic configuration. 
The threefold orbital degeneracy of this configuration accounts for a description with an effective angular momentum $L=1$~\cite{para_resonance}.  
The unquenched angular momentum in this compound yields a splitting of effective angular momentum $J=3/2$ and $J=1/2$ under the spin-orbital coupling (SOC),
and the ground state is a $J=1/2$ doublet. 
In contrast to the general belief that SOC is less important in $3d$ transition metal systems, cobalt compounds such as CoO, KCoF$_{3}$, CoCl$_{2}$ etc. feature the unquenched orbital angular momentum, leading to the pronounced SOC effect on the ground state and the macroscopic magnetic response~\cite{kCoF3_Buyers_1971, kCoF3_Holden_1971, CoO_1974, cscoCl3_1995, Co_Kitaev_model_prb2018, Co_Kitaev_model_prl2020, Co_Liu_Huimei}. 
Na$_{2}$BaCo(PO$_{4}$)$_{2}$, on the other hand, is another promising system contributing to this category. 
The experimentally estimated $g$-factors significantly deviate from 2 for both in-plane and out-plane components~\cite{Na2BaCo(PO4)2_liwei_2022,Co21128_huang2022thermal,Na2BaCo(PO4)2_mei_2022}, signaling the non-negligible orbital contribution.  
However, in the other two systems, i.e. Na$_{2}$BaX(PO$_{4}$)$_{2}$ (X=Mn, Ni),  the orbital angular momentum is quenched. 
Both the in-plane and out-plane $g$-factor is close to that of free spin~\cite{prb_sun2021_Na21128,Mn21128_2022}. 

Despite the stark difference in the active orbital angular momentum in these systems, their overall phase diagram and the QSSTs are highly similar to each other. 
Na$_2$BaMn(PO$_4$)$_2$~\cite{Mn21128_2022},  Na$_2$BaCo(PO$_4$)$_2$~\cite{Na2BaCo(PO4)2_li_possible_2020,Na2BaCo(PO4)2_mei_2022,Na2BaCo(PO4)2_liwei_2022}, and Na$_2$BaNi(PO$_4$)$_2$~\cite{cpb_Ding_2021, prb_sun2021_Na21128} display similar behavior as a function of temperature and external magnetic field. 
Under a small external magnetic field, various spin state transitions occur indicating a strong easy-axis anisotropy of these systems. 
A clear 1/3 plateau appears in the magnetization curve at low temperatures strongly suggesting the quantum nature of the magnetic coupling. 
The magnetic specific heat shows a sharp peak at T$_N$ at zero field, and the peak position shifts to other temperatures upon the application of a small external field, suggesting these systems are highly tuneable among different spin states. 
The magnetic entropy below T$_N$ only accounts for a small portion of the full entropy $R \ln 3$, indicating the existence of strong spin fluctuations above the long-range magnetic order temperature T$_N$. 
Despite these three systems having clearly different spin quantum numbers, the field dependence of the magnetic specific heat, the susceptibility, and the QSSTs all show great similarity, which is not clearly understood so far.

By using first-principle calculations and advanced many-body methods, we studied the magnetic and electronic response of Na$_2$BaX(PO$_4$)$_2$ (X = Mn, Co, Ni). 
We found that their similar crystal structures lead to a very similar electronic structure of the three systems. 
Both show correlated flat bands around the Fermi level. 
However, the trigonal crystal field and the different valence $d$ electron numbers in the three compounds lead to different electron configurations. 
Half-filled flat $e_{g}$ bands are observed in Na$_2$BaNi(PO$_4$)$_2$, which are only a quarter filled in Na$_2$BaCo(PO$_4$)$_2$. 
In Na$_2$BaMn(PO$_4$)$_2$, the flat bands at the Fermi level are $t_{2g}$ three-fold bands which are $5/6$-filled.
Despite the different electron fillings, small onsite Coulomb repulsions can always drive these systems to Mott insulators, manifesting them as ideal correlated flat band systems. 
To understand the field-driven QSSTs, we studied different collinear antiferromagnetic configurations and found that the energy difference between them is very small, which implies that they all have multiple near-degenerate magnetically states competing with the ground states, which is consistent with the experiment \cite{cpb_Ding_2021,prb_sun2021_Na21128}. 
In addition, our random phase approximation (RPA) susceptibilities show that the canted $120^{\circ}$ antiferromagnetic order becomes robust with increasing the on-site Coulomb interaction, which is also highly consistent with the experimental reports \cite{cpb_Ding_2021,prb_sun2021_Na21128}.
\begin{figure*}
\includegraphics[width=1.0\textwidth]{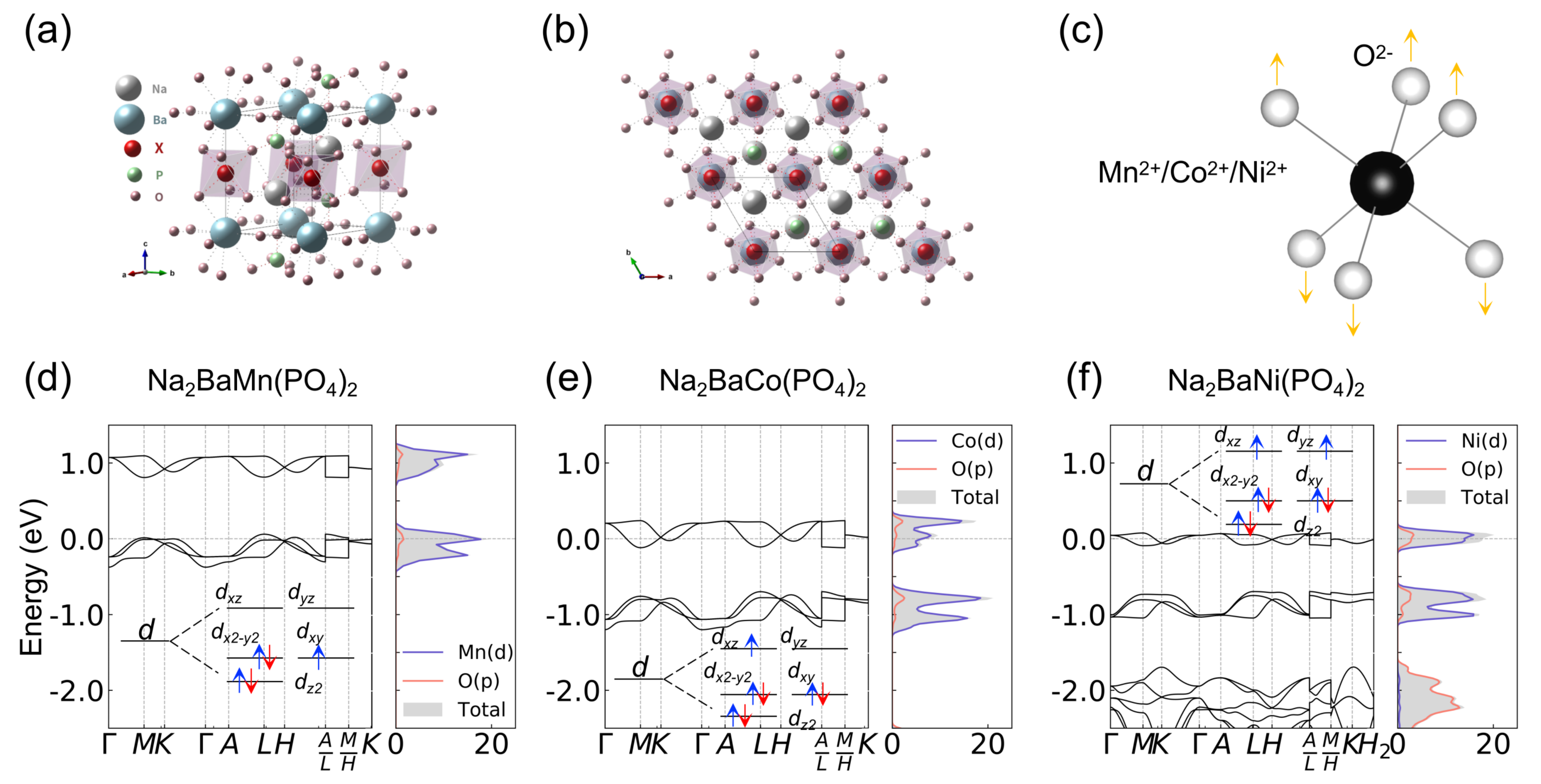}
\caption{\label{electronic_structures} The crystal structure and the DFT electronic structures of Na$_2$BaX(PO$_4$)$_2$ (X = Mn, Co, Ni).
	(a, b) The side and top view of Na$_2$BaX(PO$_4$)$_2$ crystal structure. 
    NiO$_{6}$ octahedron layer stacks along the $c$-axis. Inside the ab-plane, NiO$_6$ forms an isotropic triangular lattice without sharing an edge or corner. 
	(c) The trigonal distortion of XO$_6$ octahedron leads to the splitting of $d$-orbitals into $e_g$ and $t_{2g}$ orbitals. 
    The yellow arrow indicates the trigonal distortion of the octahedron. 
	(d - f) The electronic structures of Na$_2$BaX(PO$_4$)$_2$ (X = Mn, Co, Ni) display two blocks of flat bands around the Fermi level, corresponding to the two-fold and three-fold multiplets. 
    These flat bands mainly consist of the X-$d$(blue) orbitals as shown in the density of states plot. 
    The inset of each plot shows the electron configuration inspected from the DFT calculations.}
\end{figure*}

\section{RESULTS}

\subsection{Electronic structure of the high-temperature nonmagnetic phase}

Na$_2$BaNi(PO$_2$)$_2$ crystallizes in space group $P \bar 3$, which contains only two nontrivial generators, i.e., a threefold rotation along c-axis $\hat{c}_{3z}$ and an inversion $\mathcal{I}$. 
Na$_2$BaCo(PO$_2$)$_2$ and Na$_2$BaMn(PO$_2$)$_2$ share a similar structure but with an additional 2-fold rotational symmetry $ \hat{c}_{2[110]} $ along [110]-axis. 
They both crystallize in space group $P\bar{3}m1$. 
In the high-temperature paramagnetic phase, which we study in this work, the simultaneous presence of the inversion $\mathcal{I}$ and time-reversal $\mathcal{T}$ symmetry leads to a Kramer’s degeneracy in the Brillouin zone (BZ). 

Every band is doubly generated with its Kramer’s pair. 
XO$_6$ octahedron in Na$_2$BaX(PO$_2$)$_2$ constitute triangular lattice layer in the $ab$-plane (see Fig.~\ref{electronic_structures}(a) and (b)), which are separated by nonmagnetic BaO$_{12}$ polyhedral layers. 
XO$_6$ octahedron stack in simple $AA$-pattern along the $c$-axis without stacking fault, unlike many other hexagonal-layered magnets. 
PO$_4$ tetrahedra and Na fill the interstitial region of the XO$_6$ layers. 
We note that different XO$_6$ octahedrons are geometrically independent of each other. 
Neighboring XO$_6$ does not share an edge or corner, and the closest pathway for the magnetic interactions is through X-[PO$_4$]-X, leading to a superexchange type magnetic coupling between different local moments. 
On the other hand, the hopping in the interlayer is along X-[PO$_4$]-[PO$_4$]-X pathway. 
The interlayer magnetic coupling, thus, manifests a higher superexchange form than the intralayer one.

We carried out the first-principles calculations~\cite{DFT_1, DFT_KS_equation} with the Vienna ${\it ab-initio}$ simulation package (VASP)~\cite{VASP} by using the GGA-PBE exchange-correlation functional~\cite{PBE}. 
The plane-wave cut-off energy is 600 eV. 
We used the Monkhorst-Pack scheme to sample the first BZ by setting the spacing between $k$-points to $0.03^{-1}$. 
Structural relaxation was performed until the Hellmann-Feynmann forces on each atom are less than $5\times10^{-3}$ meV/\AA  ~before calculating the electronic structure. 

Fig.~\ref{electronic_structures}(d - f) display the energy bands of Na$_2$BaX(PO$_4$)$_2$ (X = Mn, Co, Ni) and their corresponding density of states (DOS). 
In three systems, we observe isolated blocks of bands at the Fermi level and around -1.0 eV (for Ni and Co) or 1.0 eV (for Mn), which are two-fold ($e_g $) and three-fold ($t_{2g} $) multiplets, correspondingly. 
They have a very small bandwidth. 
The emergence of the flat bands in these three isotropic triangular lattices is a consequence of the special connection of XO$_6$ (X = Mn, Co, Ni) octahedron. 
These XO$_6$ (X = Mn, Co, Ni) octahedrons behave as isolated molecules embedding in the triangular network separated by PO$_4$ tetrahedra and BaO$_{12}$ polyhedra. 
From the DOS plot, it is clear that all five bands mainly consist of $d$-orbitals of nickel, cobalt, and manganese. 
These $d$-orbitals do not directly overlap at sites from different unit cell. 
They behave more like atomic energy levels. 
The small dispersion observed in Fig.~\ref{electronic_structures}(d - f) are mainly from the hybridization with oxygen atoms inside the octahedron. 
Na$_2$BaX(PO$_4$)$_2$ (X = Mn, Co, Ni) realize the desired correlated flat bands in disorder-free single crystals, which is ideal for the experiment to study the electronic and magnetic response of flat bands in pristine solids. 
These flat bands are inevitably unstable with respect to the electronic correlation that we will study in the next section.

\begin{figure*}
\includegraphics[width=1.0\textwidth]{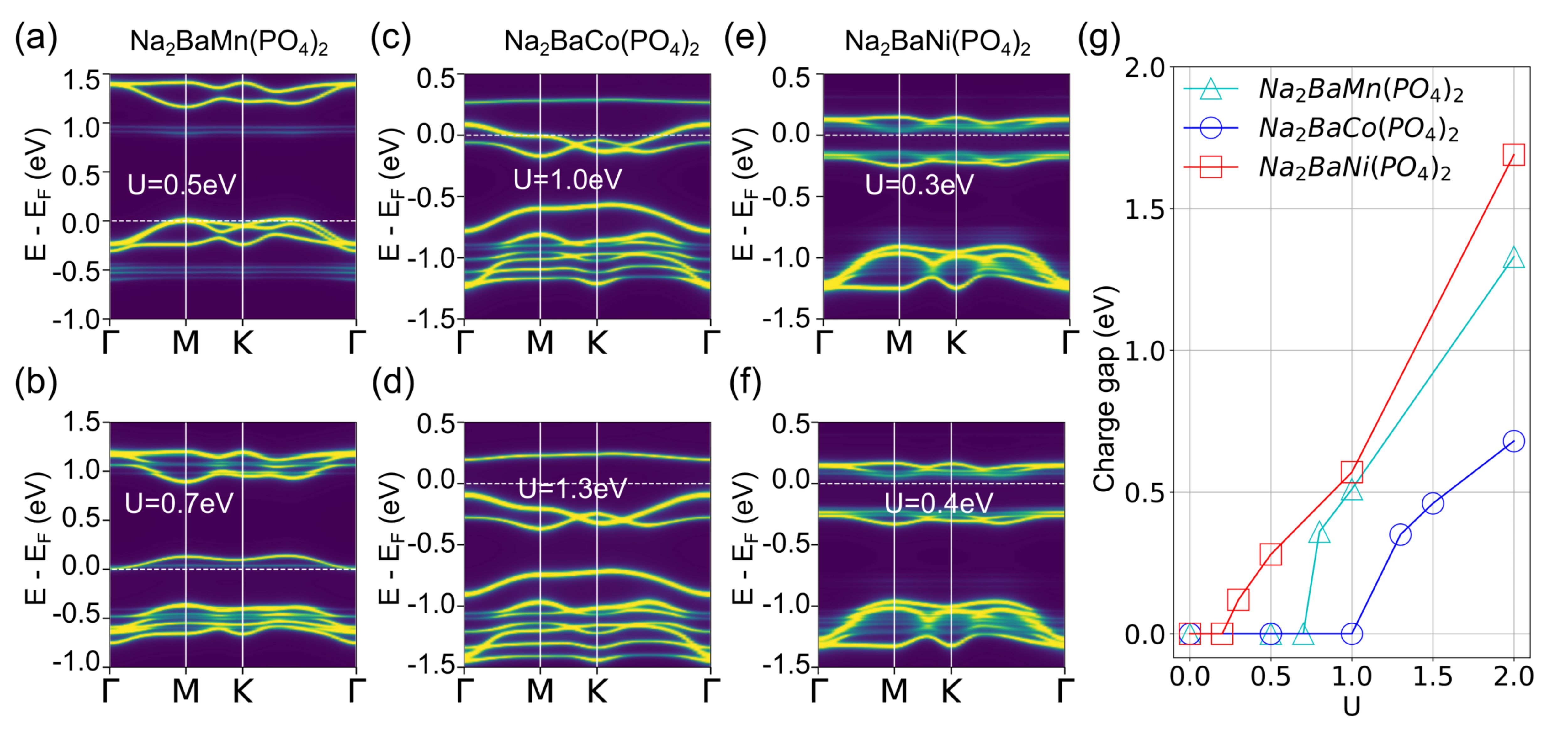}
\caption{\label{Hubbard-I} Charge instability of the flat bands driven by the local Coulomb interactions.
	(a)-(f) display the DMFT spectral functions of Na$_2$BaX(PO$_4$)$_2$ (X = Mn, Co, Ni) at T = $0.025$ eV$^{-1}$ and $J = U/10$. 
    Different $U$ values are taken to highlight the Mott transition driven by the Coulomb interactions. 
        (g) shows the evolution of the Mott gap as a function of interaction strength U.
	}
\end{figure*}

Due to the difference in the subtle symmetry and chemical environment, Na$_2$BaX(PO$_4$)$_2$ (X = Mn, Co, Ni) also display differences in their electronic structures. 
First, in Na$_2$BaNi(PO$_4$)$_2$ the $e_{g}$ bands have smaller bandwidth than those in Na$_2$BaCo(PO$_4$)$_2$ and Na$_2$BaMn(PO$_4$)$_2$. 
The same conclusion also applies to the $t_{2g} $ bands, but the difference in bandwidth is smaller as compared to the $e_g$ bands. 
This is due to the absence of 2-fold rotational symmetry $\hat{c}_{2[110]}$ in Na$_2$BaNi(PO$_4$)$_2$. 
Electrons traveling from one nickel atom to another experience a more frustrated pathway than in Na$_2$BaCo(PO$_4$)$_2$ and Na$_2$BaMn(PO$_4$)$_2$, leading to a more localized Ni-$d$ orbitals.
Second, the band filling of three systems at the Fermi level is different.
The $e_g$ bands at the Fermi level in Na$_2$BaNi(PO$_4$)$_2$ are half-filled (see Fig.~\ref{electronic_structures}(f) inset), while they are only a quarter filled in Na$_2$BaCo(PO$_4$)$_2$ (see Fig.~\ref{electronic_structures}(e) inset), and the $t_{2g}$ bands are $5/6$-filled in Na$_2$BaMn(PO$_4$)$_2$ (see Fig.~\ref{electronic_structures}(c) inset). 
XO$_6$ (X = Mn, Co, Ni) octahedron create the same local crystal field for $X^{2+}$  (X = Mn, Co, Ni). 
The octahedron is slightly distorted changing the local symmetry from $O_h$ to $D_{3d}(\bar 3m)$. 
Mn$^{2+}$, Co$^{2+}$, and Ni$^{2+}$ have $d^5$, $d^7$ and $d^8$ valence electrons yielding a different occupancy at the Fermi level of the three systems (see Fig.~\ref{electronic_structures}(c) and the inset of Fig.~\ref{electronic_structures} (d - f)). 
Thus, different responses to the electronic correlations in these three systems are expected due to the different electron occupancies. 
Third, we can roughly estimate the strength of electronic correlation acting on the flat bands from the DFT electronic structure.
The conduction bands stay at 0.81 eV in Na$_2$BaMn(PO$_4$)$_2$, 2.54 eV in Na$_2$BaCo(PO$_4$)$_2$, and 3.59 eV in Na$_2$BaNi(PO$_4$)$_2$, respectively.  
Their screening effect on the flat bands is, thus, Na$_2$BaMn(PO$_4$)$_2$ $>$ Na$_2$BaCo(PO$_4$)$_2$ $>$ Na$_2$BaNi(PO$_4$)$_2$. 
While, the highest valence bands stay at -2.75 eV in Na$_2$BaMn(PO$_4$)$_2$, -2.53 eV in Na$_2$BaCo(PO$_4$)$_2$, and -1.69 eV in Na$_2$BaNi(PO$_4$)$_2$, respectively. 
Their screening effect on the flat bands is Na$_2$BaMn(PO$_4$)$_2$ $<$ Na$_2$BaCo(PO$_4$)$_2$ $<$ Na$_2$BaNi(PO$_4$)$_2$.
Consequently, the total screening from the high-energy bands and the absolute values of electronic correlations acting on the flat bands are likely similar in these three systems. 
However, due to the difference in bandwidth, we anticipate a slightly different response of the three systems to the local electronic correlations in the paramagnetic phase.

\subsection{Paramagnetic metal-insulator transition}

\begin{figure*}
\includegraphics[width=1.0\textwidth]{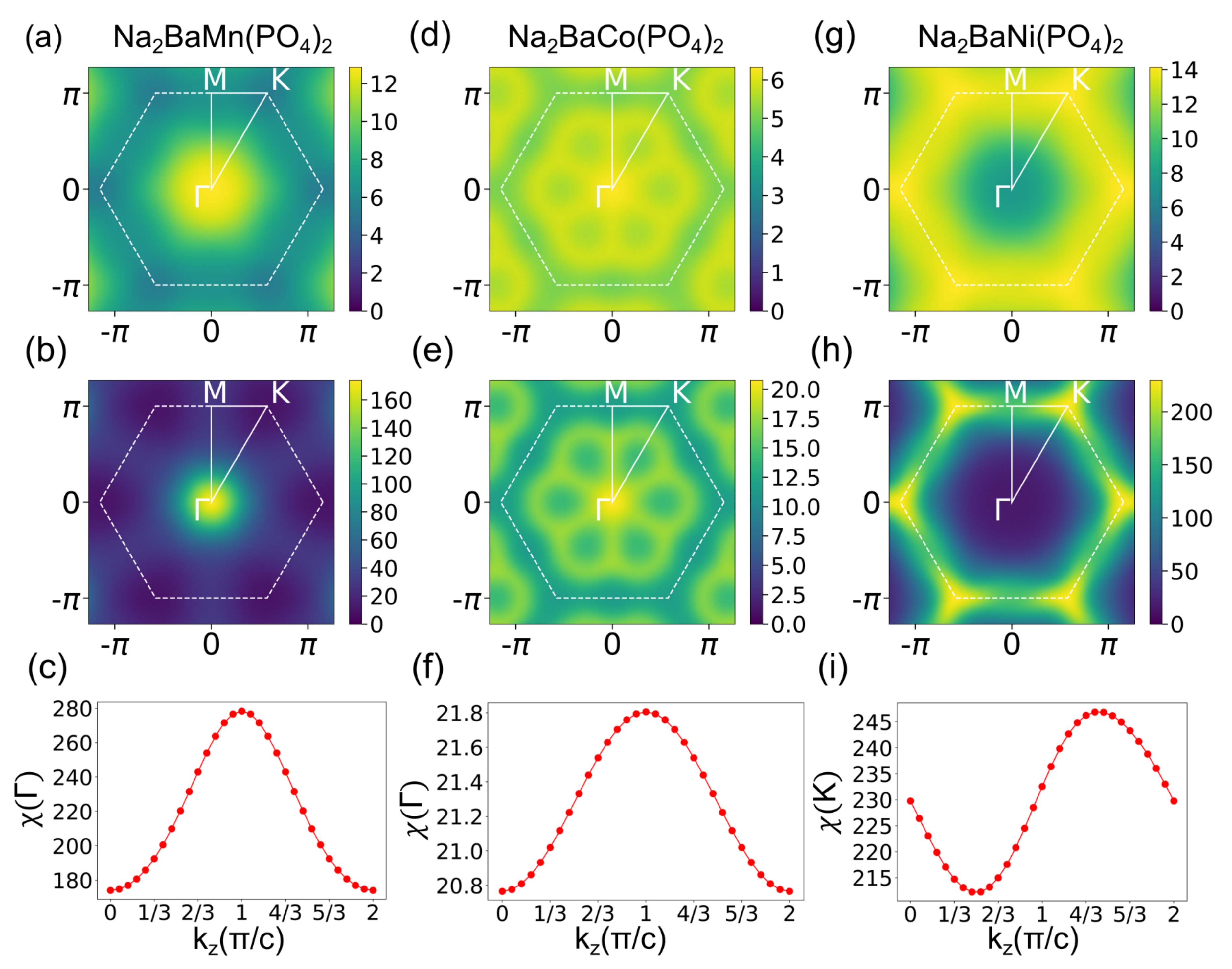}
\caption{\label{spin-sus} Spin Susceptibility.
	(a), (d) and (g) is the bare spin susceptibility of Na$_2$BaX(PO$_4$)$_2$ (X = Mn, Co, Ni) in the first BZ at $k_z = 0$. 
	(b), (e) and (h) is the RPA spin susceptibility of Na$_2$BaX(PO$_4$)$_2$ (X = Mn, Co, Ni) in the first BZ for $U \sim 0.18, 0.20, 0.12$ eV. 
	(c), (f) and (i) is the corresponding RPA spin susceptibility of the $\Gamma$ point [Na$_2$BaMn(PO$_4$)$_2$ and Na$_2$BaCo(PO$_4$)$_2$] and $K$ point [Na$_2$BaNi(PO$_4$)$_2$] in the $k_z$ direction. }
\end{figure*}

After understanding the DFT electronic structure of the three systems, we now study the electronic correlation effect on the flat bands. 
To this end, we employed dynamical mean-field theory (DMFT) \cite{DMFT_1996} to fully account for the local correlation effect with a full Kanamori interaction. 
DMFT maps the correlated lattice problem to an effective Anderson impurity model featuring the local dynamical fluctuations. 
As the effective triangular lattice formed by XO$_6$ (X = Mn, Co, Ni) octahedron is frustrated and the low-energy bands are flat, the local approximation of the DMFT is well justified. 
We first constructed a Wannier tight-binding (TB) model for all the five flat bands by using Wannier90 package~\cite{w90}. 
One-shot DMFT calculations were then performed based on this TB model. 
We set the simulation temperature $T \sim 290$ K ($\beta = 40 $ eV) and performed DMFT with a series of on-site Hubbard $U$ with fixed $J = 0.1U $. 
The quantum impurity problem was solved by the Hubbard-I solver as implemented in TRIQS packages \cite{TRIQS/DFTTools,triqs,triqs_wien2k_full_charge_SC,triqs_wien2k_interface}.

As discussed in the previous section, the local electron configurations in the three systems are different. 
In Na$_2$BaNi(PO$_4$)$_2$, the $t_{2g}$ multiplets are fully occupied and the $e_g$ doublet at the Fermi level is half-filled. 
In Na$_2$BaCo(PO$_4$)$_2$, the $t_{2g}$ states are fully occupied, while the $e_{g}$ states are 1/4 filled. 
In Na$_2$BaMn(PO$_4$)$_2$, the Mn-$d$ $e_g$ states are completely empty and the $t_{2g}$ states are 5/6 filled. 
Despite the different electron filling of these correlated flat bands, they all can be driven into the Mott insulating phase.
It is well known that the multi-orbital Hubbard model with full Kanamori interaction displays several metal-insulator transitions under electronic correlations. 
At all commensurate fillings, electronic correlations can drive the Mott transition, but the required correlation strength can be different. 
The phase diagram is symmetric with respect to half-filling. 
The Mott transition occurs at half-filling with the smallest critical interaction strength among all the commensurate fillings. 
At other commensurate fillings, the critical interaction strength gradually increases resulting in a Janus-face phase diagram. 
At incommensurate fillings or before the Mott transitions at commensurate filling, there are large phase spaces characterized as Hund's metals, which are metallic states with very low quasiparticle weight. 
Hund's rule coupling is believed to be the key concept for the understanding of iron superconductors~\cite{Hund_couple_PhysRevLett2011, Hund_couple}. 

The paramagnetic phase of these flat band systems can be viewed as multi-orbital Hubbard models with Hund's coupling at different electron fillings. 
Na$_2$BaMn(PO$_4$)$_2$, Na$_2$BaCo(PO$_4$)$_2$, and Na$_2$BaNi(PO$_4$)$_2$ correspond to a 5/6-filled $3$-orbital model, a 1/4 filled $2$-orbital model, and a half-filled $2$-orbital, respectively. 
Thus, it is natural to expect a smaller critical interaction $U_c$ in Na$_2$BaNi(PO$_4$)$_2$ than in Na$_2$BaMn(PO$_4$)$_2$ and Na$_2$BaCo(PO$_4$)$_2$. 
Fig.~\ref{Hubbard-I}(a-b), (c-d), and (e-f) display the spectral function of Na$_2$BaMn(PO$_4$)$_2$, Na$_2$BaNi(PO$_4$)$_2$, and Na$_2$BaCo(PO$_4$)$_2$ at different correlation strengths. 
For each compound, the two choices of the Coulomb interaction parameters in Fig.~\ref{Hubbard-I} are used to highlight the opening of the Mott gap. Thus, one is smaller than $U_{c}$, while the other one is larger than $U_{c}$. 
Fig.~\ref{Hubbard-I}(g) summarizes the critical interaction $U_c$ for the Mott transition in the three systems. 
The gap size $\Delta$ is shown as a function of $U$ with fixed Hund’s coupling $J = U/10$ at $T = 290$ K. 
In the three systems, the Mott gap is linearly proportional to the Coulomb repulsion and the critical values are consistent with the Janus phase diagram of a multi-orbital Hubbard model with Hund’s coupling displaying a critical difference to the electronic correlations.
Despite of the distinct ground state quantum numbers in these systems, their high-temperature paramagnetic phases are consistent with the continuous change of their valence electron numbers from Mn to Ni, which are fully described by the Janus phase diagram of the multi-orbital Hubbard model with Hund's coupling. 
Before closing this section, we want to note that above the magnetic transition temperature $T_N$ there exist strong short-range magnetic fluctuations in three systems, which were not included in our DMFT study. 
The short-range magnetic fluctuations are diﬀicult to consider as it does not create a static order parameter. 
DMFT includes the interaction-induced local magnetic moment fluctuations, while it neglects all non-local correlations. 
The short-range magnetic fluctuations, on the other hand, can accelerate the opening of the charge gap.
To go beyond the DMFT study, one will have to employ a large cell fully incorporating the short-range magnetic order. 
Reducing temperature gradually to $T_{N}$, the length of the magnetic correlations quickly increases, and the paramagnetic phase transition driven by the Coulomb repulsion described by the DMFT becomes less important.  
Instead,  the short-range magnetic correlations will affect the electron motion in the same way as the long-range order at temperatures slightly above $T_N $. 
Thus, in the paramagnetic phase, the short-range magnetic fluctuations would also split the flat bands and trigger the metal-insulator transition. 
The actual gap measured in the experiments is always a joint consequence of both electronic correlation and magnetic fluctuations.

\subsection{Spontaneous symmetry breaking}

After understanding the origin of the flat bands and the realization of Hund's physics of the multiorbital Hubbard model in Na$_2$BaX(PO$_4$)$_2$ (X = Mn, Co, Ni), we continue to study the magnetic response of these flat bands. 
By using the multi-orbital random phase approximation (RPA), we will probe the spontaneous symmetry-breaking modes of the flat bands.

In the DMFT study, we suppressed the magnetic order by imposing the paramagnetic condition during the calculations. 
In this section, we will still concentrate on the paramagnetic phase. 
But the logic here is to look for the breaking of the paramagnetic solution, which will give us strong hints on the possible magnetic long-range order. 
This is achieved by calculating the multi-orbital spin susceptibility and examine the wave vectors where the susceptibility becomes diverging. 
We first evaluate the bare particle-hole bubble susceptibility from single-particle Green’s function defined in the Wannier TB basis. 
\begin{equation}
	\chi^0_{l_1 l_2; l_3 l_4}(\mathbf{q},i\omega_n) = - \frac{1}{\beta N_{\mathbf{k}}} \sum_{\mathbf{k},\nu_n} G^0_{l_4 l_1}(\mathbf{k}+\mathbf{q}, i\omega_n + i\nu_n) G^0_{l_2 l_3}(\mathbf{k}, i\nu_n)
\label{chi_0}
\end{equation}
Here the subscripts $l_1 - l_4$ denote the orbital indices. $\beta = 100$ eV$^{-1}$ is the inverse temperature and $N_{\mathbf{k}} = 42 \times 42 \times 30 $ is the number of $\mathbf{k}$-points used for BZ summation.
$G^0_{12}(\mathbf{k}, i\omega_n) = \left( i\omega_n - \mathbf{H}^0 \right)^{-1}_{12}$ is the non-interacting Green's function, where $\mathbf{H}^0$ is the non-interacting Hamiltonian matrix in the Wannier TB basis.
Then, we calculate the RPA spin susceptibility as
\begin{eqnarray}
\chi^{\text{RPA}}_{s}(\mathbf{q}) = [1- U^s \chi^0 (\mathbf{q})]^{-1} \chi^0(\mathbf{q}) \;,
\label{chi_spin}
\end{eqnarray}
where the interaction matrix in the spin channel takes the following standard form~\cite{RPA_multiorbital}
\begin{equation}
	U^s_{l_1 l_2; l_3 l_4} =\left\{
	\begin{array}{ll}
		U & ,l_1=l_2=l_3=l_4 \\
		J & ,l_1=l_2 \neq l_3=l_4 \\
		U^{\prime} & ,l_1=l_4 \neq l_3=l_2 \\
		J^{\prime} & ,l_1=l_3 \neq l_2=l_4\;.
	\end{array} \right.
\end{equation}
Now, the bare and RPA spin susceptibility can be evaluated by $\chi^{0/\text{RPA}}_{s}({\mathbf{Q}})
= \sum_{12} \chi^{0/\text{RPA}}_{11;22}(\mathbf{Q},i\omega_n=0)$.

In Fig.~\ref{spin-sus}, we show magnetic susceptibilities at $k_z = 0$ for Na$_2$BaMn(PO$_4$)$_2$ [Fig.~\ref{spin-sus}(a, b)], Na$_2$BaCo(PO$_4$)$_2$ [Fig.~\ref{spin-sus}(d, e)] and Na$_2$BaNi(PO$_4$)$_2$ [Fig.~\ref{spin-sus}(g, h)], respectively. 
Fig.~\ref{spin-sus}(a, d, h) display the bare magnetic susceptibility evaluated with Eq.~(\ref{chi_0}). 
It shows a continuum along the BZ boundary in Na$_2$BaNi(PO$_4$)$_2$. But the continuum is around $\Gamma$-point in Na$_2$BaMn(PO$_4$)$_2$ and Na$_2$BaCo(PO$_4$)$_2$. 
The absence of peak structure in magnetic susceptibility indicates the stability of the paramagnetic solution in the non-interacting limit.
However, after including the electronic correlations, the RPA magnetic susceptibility calculated from Eq.~(\ref{chi_spin}) clearly develops a profile at the K-point in Na$_2$BaNi(PO$_4$)$_2$, and at the $\Gamma$-point in Na$_2$BaMn(PO$_4$)$_2$ and Na$_2$BaCo(PO$_4$)$_2$, which indicates to a two-dimensional $120^{\circ}$-AFM correlation of the Ni $d$-orbitals and a FM correlation of the Mn/Co $d$-orbitals. 
The different behavior of magnetic susceptibility reveals that different magnetic correlations are developed in these systems. 
The in-plane magnetic correlations in Na$_2$BaMn(PO$_4$)$_2$ are predominantly ferromagnetic in nature [see Fig.~\ref{spin-sus}(b)], while in Na$_2$BaCo(PO$_4$)$_2$, the in-plane magnetic correlations are ferromagnetic, but with competition from antiferromagnetic interactions [see Fig.~\ref{spin-sus}(e)]. 
In contrast, the in-plane magnetic correlations in Na$_2$BaNi(PO$_4$)$_2$ are antiferromagnetic, with each local moment pointing to each other by $120^{\circ}$ angle [see Fig.~\ref{spin-sus}(h)].

To investigate the out-of-plane magnetic correlations, we calculated the RPA spin susceptibility $\chi _m(k)$ at different $k_z$ as a function of $k_x$ and $k_y$. 
We found that $\chi _m(k)$ of Na$_2$BaMn(PO$_4$)$_2$ and Na$_2$BaCo(PO$_4$)$_2$ show peaks at $k_x = k_y = 0 $. 
While, the peaks of $\chi _m(k)$ are at $k_x = k_y = 1/3$ in Na$_2$BaNi(PO$_4$)$_2$ for all $k_z$. 
We plot the peak value of $\chi _m[k = (0, 0, k_z)]$ for Na$_2$BaMn(PO$_4$)$_2$ and Na$_2$BaCo(PO$_4$)$_2$, and $\chi _m[k = (1/3, 1/3, k_z)]$ for Na$_2$BaNi(PO$_4$)$_2$ at different $k_z$ in Fig.~\ref{spin-sus}(c, f, i). 
As one can see, as a function of $k_{z}$, $\chi _m[k = (0, 0, k_z)]$ peaks at $k_z \sim \pi$ in Na$_2$BaMn(PO$_4$)$_2$ and Na$_2$BaCo(PO$_4$)$_2$, while the peak of $\chi _m[k = (1/3, 1/3, k_z)]$ is at $k_z \sim 3\pi/2$ in  Na$_2$BaNi(PO$_4$)$_2$. 
The non-zero value of $k_z$ indicates an out-plane antiferromagnetic spin coupling in all three compounds. 

As the variation of the peak values at different $k_z$ is small, consistent with experiments, the dominant magnetic correlation is in $ab$-plane in all three systems. 
In Na$_2$BaNi(PO$_4$)$_2$, both the in-plane 120$^{\circ}$-AFM and the out-plane AFM couplings inspected in $\chi_{m}$ calculations are consistent with experimental estimation~\cite{prb_sun2021_Na21128}.
However, the RPA results of $\chi_{m}$ in Na$_2$BaMn(PO$_4$)$_2$ and Na$_2$BaCo(PO$_4$)$_2$ systems are inconsistent with experimental observations \cite{Mn21128_2022, Na2BaCo(PO4)2_mei_2022, Na2BaCo(PO4)2_liwei_2022, Na2BaCo(PO4)2_li_possible_2020, Na2BaCo(PO4)2_zhong_strong_2019}, which also indicate an in-plane AFM magnetic coupling. 
The failure of $\chi_{m}$ in predicting the nature of exchange coupling for these two systems is likely due to the missing of the superexchange process between transition metal ions and oxygen, which was not included in the downfolded tight-binding model with only $d$-orbitals for spin susceptibility calculations.
It indicates that, unlike Na$_2$BaNi(PO$_4$)$_2$, the effective spin exchange couplings are more sensitive to the hybridization with oxygen in Na$_2$BaMn(PO$_4$)$_2$ and Na$_2$BaCo(PO$_4$)$_2$. 

Going beyond the qualitative understanding of the magnetic coupling, we further consider a tight-binding model with 5 $d$-orbitals of the transition metal ion and 24 $p$-orbitals of the oxygen. Additionally, we also included the SOC in the DFT calculations. 
Based on this $d-p$ model, we ${\it ab-initio}$ determine the exchange couplings by employing the TB2J~\cite{TB2J} package, which maps the DFT ground state to a general spin model with isotropic Heisenberg term, off-diagonal anisotropic term, as well as the Dzyaloshinskii-Moriya interaction, and the single-ion anisotropy term. 
All parameters can be estimated from spin-polarized DFT calculations. 
By performing a ferromagnetic calculation, we obtained the antiferromagnetic exchange constants $(J_{xx}, J_{yy}, J_{zz})$ of Na$_2$BaMn(PO$_4$)$_2$ and Na$_2$BaCo(PO$_4$)$_2$ are $(0.52 K, 0.52 K, 0.57 K)$ and $(0.87 K, 0.91 K, 1.05 K)$, respectively. 
These values are in accordance with other estimations~\cite{Mn21128_2022, Na2BaCo(PO4)2_mei_2022, Na2BaCo(PO4)2_liwei_2022, Na2BaCo(PO4)2_li_possible_2020, Co21128_J_Kataev2021}, while the other off-diagonal terms are negligibly small in our calculations.
These two systems can be effectively understood as XXZ models. 
The isotropic antiferromagnetic exchange constant of Na$_2$BaNi(PO$_4$)$_2$ is $1.53 K$ and the anisotropy value is minuscule, exhibiting qualitative consistency with experimental results~\cite{prb_sun2021_Na21128} but also with a slight quantitative deviation.
We note that the effective models considered here are conceptually different from the models extracted directly from experiments, where the ground states have been restricted exactly to the ones with a presumed total quantum number. 
For example, in Na$_2$BaCo(PO$_4$)$_2$ the total quantum number is believed to be $1/2$ with contribution from the active orbital angular momentum $L=1$. 
The spin model then corresponds to the couplings of these effective spins. 
Here, by using the first-principles approach we actually obtained the interactions in the Hilbert space with real spins.
We obtained the effective spin interactions between all the $d$-orbital electrons, thus, the model parameters can be slightly different from those extracted from experiments. 
While, as long as DFT correctly describes the ground state and the ground state has a large energy separation from other excited states, the {\it ab-inito} estimation of the effective spin model would coincide with the ones phenomenologically or numerically extracted from experiments. 

\subsection{Competing magnetic states}

From the calculations of the magnetic susceptibility and exchange couplings, we have convincingly shown that the ground state of Na$_2$BaX(PO$_4$)$_2$ (X = Mn, Co, Ni) favors antiferromagnetic correlations, which is highly consistent with the experiments~\cite{Mn21128_2022,Na2BaCo(PO4)2_mei_2022,Na2BaCo(PO4)2_liwei_2022,Na2BaCo(PO4)2_zhong_strong_2019,Na2BaCo(PO4)2_li_possible_2020,prb_sun2021_Na21128,cpb_Ding_2021}. 
Another important feature of Na$_2$BaX(PO$_4$)$_2$ (X = Mn, Co, Ni) is the field-driven QSSTs. 
Under small external field $ \vec{B}//a$ and $\vec{B}//c$, a rich phase diagram containing several competing magnetic states was experimentally extracted for the three systems. 
The ability of Na$_2$BaX(PO$_4$)$_2$ (X = Mn, Co, Ni) to be tuned by a small magnetic field strongly suggests that these magnetic states are highly competing in energy with each other \cite{prb_sun2021_Na21128, Na2BaCo(PO4)2_li_possible_2020, Na2BaCo(PO4)2_mei_2022, Na2BaCo(PO4)2_liwei_2022, Mn21128_2022}. 
They stay as the spin excitation states and their relative energy to the ground state must be very small. 
To further shed light on the QSSTs, we theoretically study a number of magnetic states and compare their total energies.

For simplicity, here we only consider collinear magnetic structures. 
We generate 12 different collinear magnetic structures according to the crystal symmetry and layer information by utilizing the enumeration algorithm implemented in~\cite{HT_horton_HTP_2019, HT_Su_HTP_2022}. 
To account for the electronic correlations, we performed the DFT + U calculations with Coulomb parameters $U = 4.710$ eV, $J = 0.575$ eV for manganese, $U = 5.237$ eV, $J = 0.807$ eV for cobalt, and $U = 5.847$ eV, $J = 0.589$ eV for nickel~\cite{DFT+U}, respectively. 
Each magnetic structure was fully relaxed before the total energy and the electronic structure were calculated. 
The appendix Fig.~\ref{FigA:energy} displays the electronic structure of the twelve magnetic configurations. 

Two conclusions can be clearly derived from the calculation: (1) The energy difference between different collinear magnetic structures is very small. 
(2) The electronic structures look very similar. 
From the DOS plot, we notice that the valence band top is dominated by the O-$p$ states. 
The X-$d$(X = Mn, Co, Ni) orbitals now stay at $\pm2.5$ eV. 
The conduction bands of X-$d$(X = Mn, Co, Ni) states remain flat but the valence bands gain significant bandwidth. 
This general feature is observed in all magnetic structure calculations, which is attributed to the locality of the XO$_6$ (X = Mn, Co, Ni) octahedra.

As we have learned in Fig.~\ref{electronic_structures}, the flat bands of X-$d$(X = Mn, Co, Ni) orbitals originate from the large lattice-spacing of the XO$_6$ (X = Mn, Co, Ni) octahedra from each other. 
They can be safely taken as small molecules embedded in the crystal. 
The absence of direct $d$-$d$ hopping results in a small bandwidth of these $d$ bands. 
Starting from this nonmagnetic state, the spin polarization in each magnetic structure locally acts on the $d$-orbitals as a Zeeman field, which splits the flat bands at the Fermi level and pushes them to the high energy. 
Some flat bands merge into the valence leading to a strong hybridization with the O-$p$ states, thus, they gain significant bandwidth. 
Some flat bands in the conduction remain isolated from the other conduction bands, thus, their bandwidth does not increase.

The small energy difference and the response of the flat bands to the spin polarization are consequences of the local nature of the XO$_6$ (X = Mn, Co, Ni) octahedra. 
The low-energy excitations of the three systems are dominated by the isolated octahedra manifesting atomic-like energy levels. 
The magnetic couplings of those $d$-orbitals are of superexchange type that is frustrated by the triangular network. 
Thus, different magnetic states have very similar total energy and electronic structures. 
Such small differences can be easily suppressed by quantum fluctuations leading to QSL states.

\section{Conclusion}

In summary, we have theoretically studied three triangular TM compounds Na$_2$BaX(PO$_4$)$_2$ (X = Mn, Co, Ni) and discovered correlated flat bands in the three systems. 
These flat bands nicely follow the trigonal field splitting and demonstrate atomic-like electron occupancy and energy levels. 
The presence of flat bands in the three systems is a consequence of the spatially isolated XO$_6$ (X = Mn, Co, Ni) octahedra. 
The absence of direct $d$-$d$ electron hopping features the three systems of the frustrated superexchange couplings between local moments. 
By using first-principles and advanced many-body methods, we studied the instability of the flat bands against electronic and magnetic correlations. 
The different electron fillings in Na$_2$BaX(PO$_4$)$_2$ (X = Mn, Co, Ni) nicely realize a multi-orbital Hubbard model with Hund’s coupling. 
The three systems nicely resemble the different insulating phases in the Janus face phase diagram.

Without imposing constraints on the final state, our spontaneous symmetry-breaking calculations of the magnetic susceptibility and exchange constant clearly indicate an antiferromagnetic coupling in the ab-plane and antiferromagnetic coupling along the $c$-axis in Na$_2$BaX(PO$_4$)$_2$ (X = Mn, Co, Ni), which qualitatively agrees with the experiment.
In the three systems, the magnetic susceptibility at different $\mathbf{q}$ shows an overall similar amplitude indicating a strong competing nature of various types of magnetic correlations. 
We confirmed this conclusion by explicitly exploring a few collinear magnetic structures and found that they all have similar total energy and electronic structures. 
The latter can be easily understood from the picture of isolated XO$_6$ (X = Mn, Co, Ni) octahedra. 
The presence of local moments and the breaking of time-reversal symmetry act as a local Zeeman field on the flat bands in the three systems. 
We conclude that Na$_2$BaX(PO$_4$)$_2$ (X = Mn, Co, Ni) are dominated by the atomic-like low-energy excitations originating from the large spatial separation of XO$_6$ (X = Mn, Co, Ni) octahedra. 
Their critical difference in these paramagnetic states lies in the different electron occupancies, which result in a different response to the electronic correlations and spin polarizations. 
Our work unifies the three structurally similar systems but with clearly different spin quantum numbers, and paves the way for further exploring correlated flat bands in QSL states.

\begin{acknowledgments}
This work was supported by the National Key R\&D Program of China under Grant No. 2022YFA1402703, the National Natural Science Foundation of China (11874263), 2021-Fundamental Research Area 21JC1404700, Shanghai Technology Innovation Action Plan (2020-Integrated Circuit Technology Support Program 20DZ1100605), and Sino-German Mobility program (M-0006). 
X.F.Z. also acknowledges the support from the Postdoctoral Special Funds for Theoretical Physics of the National Natural Science Foundation of China (12147124). 
Part of the calculations were performed at the HPC Platform of ShanghaiTech University Library and Information Services and the School of Physical Science and Technology.
\end{acknowledgments}

\appendix
\section{Competing magnetic states}
\begin{figure*}[htbp]
\centering
\includegraphics[width=\linewidth]{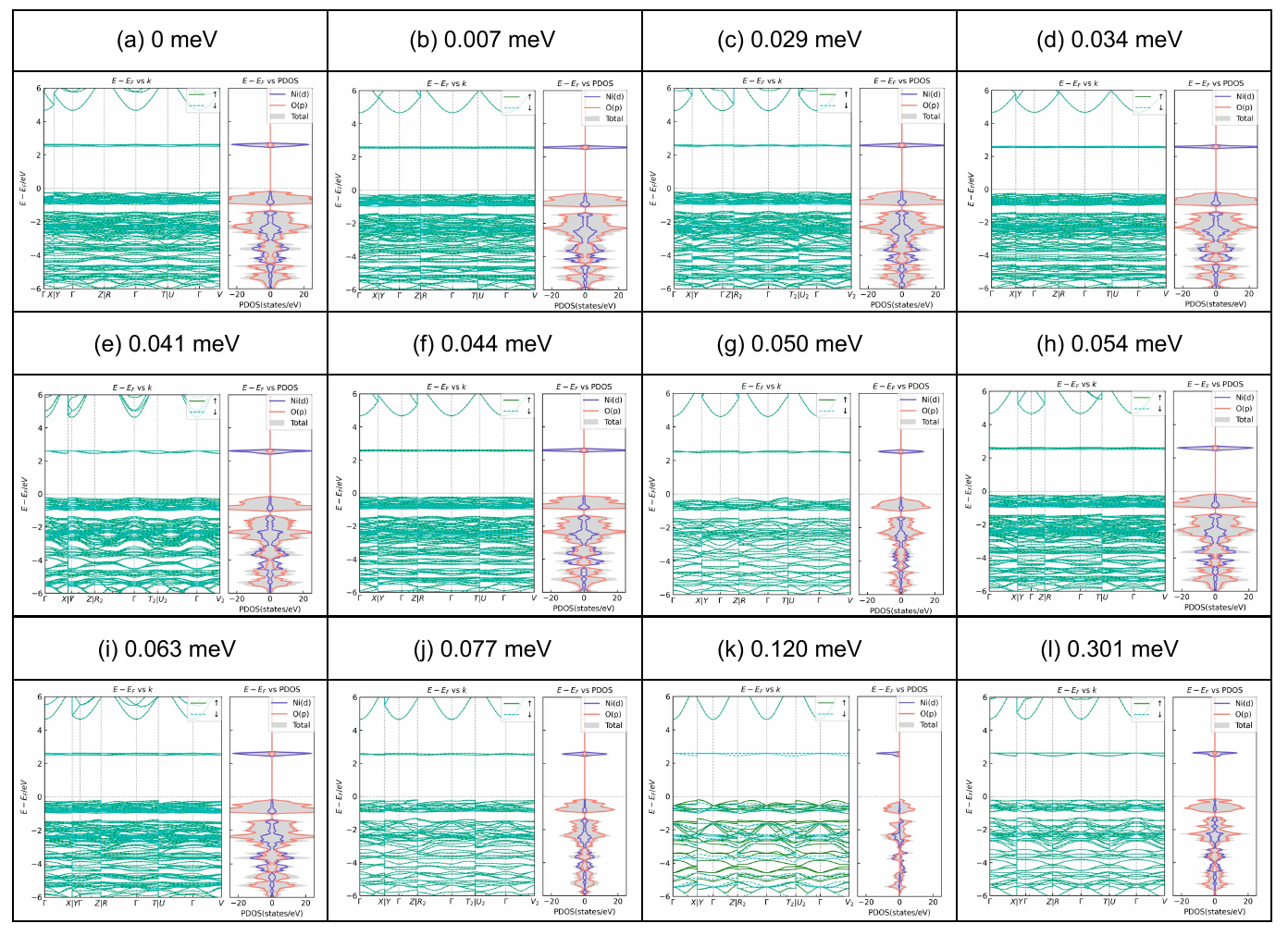}
\caption{The comparison of electronic structures of the twelve collinear magnetic configurations. The total energy of each configuration relative to the state with the lowest energy is shown in each subplot. }
\label{FigA:energy}
\end{figure*}
In this appendix, we present detailed information on the crystal and electronic structures of the 12 collinear magnetic states that we considered in the main text.  
To highlight the influence of magnetic order in the electronic structure, we performed the LSDA + U calculations without SOC, such that the spin-up and spin-down states are separately resolved. 
Here, we only list the corresponding results for Na$_2$BaNi(PO$_4$)$_2$. The conclusion also applies to Na$_2$BaMn(PO$_4$)$_2$ and Na$_2$BaCo(PO$_4$)$_2$. 
The crystal structures of these candidate states are determined automatically by the numeration algorithm~\cite{HT_horton_HTP_2019, HT_Su_HTP_2022} that utilizes the crystal symmetry and layer information of the unpolarized primitive cell. 
We include all twelve magnetic structures as a tar package in the supplement material and present the electronic structures and the corresponding DOS in the appendix Fig.~\ref{FigA:energy}. 
The twelve magnetic structures include one ferromagnetic structure and eleven antiferromagnetic structures. 
The one with the lowest total energy is antiferromagnetic, whose electronic structure and DOS are shown in Fig.~\ref{FigA:energy} (a). 
The energy differences per atom of the other structures with respect to it are shown in each subplot in units of meV/atom. 
As one can see the largest energy difference is only 0.3 meV/atom, indicating all these structures degenerate in energy. 
The small energy difference can be easily compensated by the external magnetic field, leading to the transitions between different magnetic configurations as observed in experiments. 
We also observe that the electronic structures of the different magnetic configurations are similar, including the $d$ states at around 2.5 eV and the separation of the valence bands at -1.5 eV. 
The high similarity of the electronic structures of the different magnetic configurations indicates that the magnetic long-range order is less important but the local magnetic moment fluctuations are essential. 
The magnetic couplings of the local moment are weak which is consistent with the {\it ab-initio} estimation of the exchange couplings discussed in the main text.

\bibliographystyle{apsrev4-2}
\bibliography{ref}

\end{document}